# Fully Resolved Numerical Simulations of Fused Deposition Modeling. Part I—Fluid Flow

Huanxiong Xia[1], Jiacai Lu[1], Sadegh Dabiri[2], and Gretar Tryggvason[1]

[1]Department of Aerospace and Mechanical Engineering
University of Notre Dame, IN, USA
[2]School of Mechanical Engineering and Department of Agricultural and Biological Engineering
Purdue University, West Lafayette, IN, USA.

## Abstract

**Purpose** — This paper presents a first step toward developing a comprehensive methodology for fully resolved numerical simulations of fusion deposition modeling.
**Design/methodology/approach** — A front-tracking/finite volume method previously developed for simulations of multiphase flows is extended to model the injection of hot polymer and its cooling down.
**Findings** — The accuracy and convergence properties of the new method are tested by grid refinement and the method is shown to produce convergent solutions for the shape of the filament, the temperature distribution, contact area and reheat region when new filaments are deposited on top of previously laid down filaments.
**Research limitations/implications** — The present paper focuses on modeling the fluid flow and the cooling. The modeling of solidification, volume changes and residual stresses will be described in Part II.
**Practical implications** — The ability to carry our fully resolved numerical simulations of the fusion deposition process is expected to help explore new deposition strategies and to provide the "ground truth" for the development of reduced order models.
**Originality/value** — The present paper is the first fully resolved simulation of the deposition in fusion filament modeling.

## 1   Introduction

Of the many processes for rapid prototyping collectively referred to as 3D printing, Fused Deposition Modeling (FDM), or Fused Filament Fabrication (FFF), is perhaps the best known one. In FDM complex three-dimensional objects are built by depositing filaments of hot polymers that fuse together when they solidify. FDM was patented in 1992 (Crump, 1992) and 3D printers based on FDM have been manufactured by the Stratasys corporation for several years. The process is a very versatile way to generate fairly complex geometries, but the basic aspects are relatively simple. As the original and related patents have expired a large number of low cost 3D printers based on FDM have become available from a number of developers, and interest in the process has increased.

The physical processes behind FDM are easily understood but the quality of the final product depends sensitively on exactly how the process is carried out and current understanding of how to select the various control parameters has mostly been achieved by experimentation. A few examples of experimental studies include the effect of cooling rate on the bonding of filaments

(Sun et al., 2003); the anisotropy of the properties of ABS parts (Ziemian et al., 2012a; Nikzad et al., 2011); the effect of color and temperature (Wittbrodt and Pearce, 2015); dimensional tolerances and the sources of error (Bochmann et al., 2015; Lieneke et al., 2016); deposition on a curved bed (Allen and Trask, 2015); and the potential for using polypropylene for FDM parts (Carneiro et al., 2015). Experimental studies have also been done to optimize the process (Kaveh et al., 2015; Mohamed et al., 2016), including using advanced statistical techniques and machine learning (Rao and Rai, 2016). As FDM is used for more complex situations, and as demand for more reliability increases, it is likely that a better theoretical understanding would be helpful (Guessasma et al., 2015).

We believe, in particular, that the ability to conduct detailed and fully resolved numerical simulations to examine how the polymer is injected, cools down and fuses with the material already in place, will result in improved predictive capabilities. Those should facilitate better control and the ability to build more complex structures, such as materials with graded properties due to variable nozzle speed and temperature, optimization, the co-injection of many materials and adding of reinforcements, as well as printing with materials that are currently difficult to use. Fully resolved simulations can also produce input for reduced order models intended for rapid and routine modeling. Here we present a method to simulate the formation of fully three-dimensional objects by laying down hot filaments that become rigid as they cool down. The method is based on a numerical method originally developed for direct numerical simulations of multiphase flow that has already been used to examine a large number of complex flows, see Tryggvason et al., 2011 for a detailed description and Tryggvason et al., 2001; Al-Rawahi and Tryggvason, 2004, and Vu et al., 2013 for a few examples of studies using the method. Here, the necessary extensions of the method are described, detailed convergence studies for a simple geometry are presented, and an example of the construction of a more complex fully three-dimensional object shown. While fully resolved simulations should allow us to examine all aspects of the process, including the geometry, shrinkage, surface roughness, fused area, bonding strength, interior stress, processing parameters, deposition strategy and other important issues, much of the shape is determined by the dynamics of the molten polymer and the heat transfer. In the present paper we therefore focus on the fluid mechancs and heat transfer and the filaments "solidify" only because their viscosity becomes high. We also use a simple source model for the injection. The incorporation of a model for the solidification and a more complete description of the nozzle will be described in later publications. For initial efforts to simulate the deposition of filaments see Dabiri et al., 2014, where we used very "benign" material properties to make the simulations easy.

Modeling efforts for FDM, both analytical and computational, have so far focused on various parts of the process, such as the cooling down after the filament has been deposited (Ji and Zhou, 2010), mechanical properties of the finished part (Zhang and Chou, 2008; Ziemian et al., 2012b; Singamneni et al., 2012; Domingo-Espin et al., 2015; Shahrain et al., 2016; Lieneke et al., 2016), the flow in the nozzle (Ramanath et al., 2008), the bond between filaments (Bellehumeur et al., 2004), and the roughness of the finished part (Ahn et al., 2009; Boschetto and Bottini, 2015). The only other fully resolved simulation of the deposition process, that we are aware of, is Bellini, 2002, who modeled the fluid flow and heat transfer during the extraction of ceramic filaments, assuming a two-dimensional setup. For a review focused on FDM and the state-of-the-art in modeling, see Turner et al., 2014 and Turner and Gold, 2015.

Other efforts to simulate the formation of a solid object by additive manufacturing have focused on laser melting of metal particles (or metal powder bed fusion), see Markl et al., 2015, for example, and King et al., 2015, for a review of work done at the Lawrence Livermore National

Laboratory. We also note that several numerical simulations of the deposition and solidification of metal drops have been reported in the literature. Those studies were, in many cases, motivated by spray forming and plasma coating and the droplets therefore generally had high impact velocities and splatted on impact, forming a thin layer (Liu et al., 1993, 1995; Fukai et al., 1995; Chung and Rangel, 2001; Pasandideh-Fard et al., 2001, 2002; Mostaghimi et al., 2002). High impact velocities generally result in little control over the placement of individual drops but lower impact velocities offer more control and the possibility of stacking the drops to make complex objects as discussed by Gao and Sonin, 1994. That process was examined computationally by Che et al., 2004, who assumed an axisymmetric geometry and used the results to identify process parameters that resulted in "good" structures.

A recent review of additive manufacturing, including FDM, can be found in Bikas et al., 2016, for example, and a summary of a recent workshop can be found in Schwalbe, 2016. Although originally used mostly for prototyping mechanical parts during the design process, FDM is increasingly being considered for a larger range of applications, including various biological ones, such as the constructions of scaffolds for bone (Abdelaal and Darwish, 2011, for example).

## 2 Problem Setup and Numerical Method

The computational domain and the deposition of a hot liquid filament on a horizontal plate by a moving volume source are shown schematically in Figure 1. The process is embedded in a hexahedral computational domain that includes the deposited object and the nozzle. The bottom of the domain (the plate) is a rigid, no-slip adiabatic wall but the top and side walls are open and allow air to freely flow out of the domain as the polymer melt is injected. The ambient air and the plate are colder than the melt, and as the melt cools down its viscosity increases until it is essentially solid. The nozzle is modeled using a volume source and by moving the source with a given velocity along a specified path, we can control where the injected melt ends up and solidifies. By depositing a filament on top of already deposited and solidified material a complex shape can be built up.

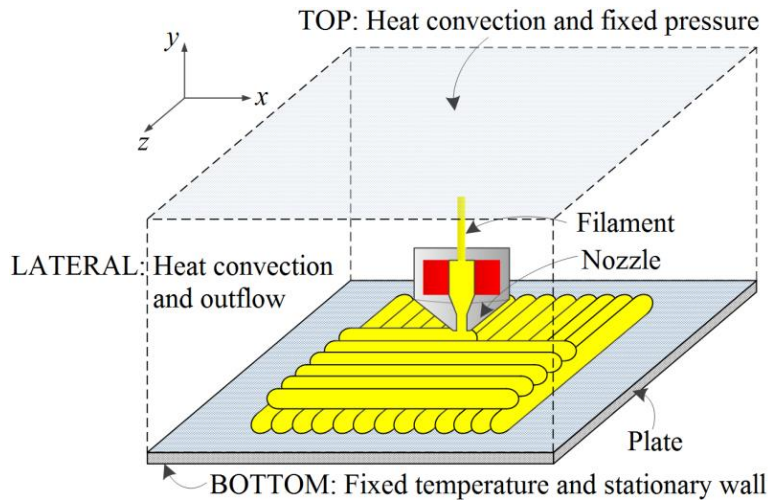

Figure 1: A schematic, showing the computational domain.

The evolution is specified by the rate of injection $\dot{Q}$, the velocity of the source $U_s$, the injection temperature $T_{inj}$, the plate temperature $T_p = 40\,°\text{C}$, and the ambient temperatures $T_\infty = 20\,°\text{C}$, along with the properties of the filament and the air, such as the density, $\rho$, the viscosity, $\mu$, the thermal diffusivity, $\alpha$, the surface tension, $\sigma$, as well as how these properties depend on the temperature. To derive the appropriate nondimensional numbers we take $U_s$ as a velocity scale and define a length scale as the diameter of the deposited filament, $D = \sqrt{4\dot{Q}/(U_s\pi)}$, assuming that it is circular. These scales give a Reynolds number, $Re = \rho DU/\mu$, and a Pécklet number, $Pe = DU_s/\alpha$, for both the melt and the air and a Weber number, $We = \rho U_s^2 D/\sigma$. For 3D printing with a PLA filament, typical processing parameters are an injection rate of $\dot{Q} = 4.81 \times 10^{-8} m^3/s$ and a nozzle velocity of $U_s = 0.10 m/s$. The values for the various properties are given in Table 1. Using those values the various nondimensional numbers can be estimated as: $Re_p = 4.9 \times 10^{-3} - 1.9 \times 10^{-5}$, $Pe_p = 995$, and $We_p = 0.243$. For the air we have $Re_a = 3.1$ and $Pe_a = 2.1$. Here the subscripts $p$ refers to the polymer and $a$ to the air.

Table 1: Material properties of the polymer and the air.

| Parameter | Value | Ref. |
|---|---|---|
| Density of the air $\rho_a(kg/m^3)$ | 0.9 | (Haynes, 2014) |
| Viscosity of the air $\mu_a(Pas)$ | $2.3 \times 10^{-5}$ | (Haynes, 2014) |
| Thermal conductivity of the air $k_a(W/mK)$ | 0.034 | (Haynes, 2014) |
| Specific heat capacity of the air $c_{p,a}(J/kgK)$ | 1000 | (Haynes, 2014) |
| Density of the polymer $\rho_p(kg/m^3)$ | 1240 | (Moldow Plastic Labs, 2007) |
| Viscosity of the polymer $\mu_p(Pas)$ (Range-function of $T$) | 20-5000 | (Moldow Plastic Labs, 2007) |
| Thermal conductivity of the polymer $k_p(W/mK)$ | 0.195 | (Moldow Plastic Labs, 2007) |
| Specific heat capacity of the polymer $c_{p,p}(J/kgK)$ | 2000 | (Moldow Plastic Labs, 2007) |
| Surface tension coefficient $\sigma(kg/s^2)$ | 0.04 | (KINO Industry Co., LTD., 2014) |

The dynamics is governed by the Navier Stokes equations. Those are in the conservative form and for the whole computational domain, including both the polymer and the air:

$$\frac{\partial \rho \mathbf{u}}{\partial t} + \nabla \cdot (\rho \mathbf{u}\mathbf{u}) = -\nabla p + \rho \mathbf{g} + \nabla \cdot \boldsymbol{\sigma} + \sigma \int_F \kappa_f \mathbf{n}_f \delta(\mathbf{x} - \mathbf{x}_f) dA_f. \quad (1)$$

Here, $\mathbf{u}$ is the velocity vector, $\rho$ and $\mu$ are the discontinuous density and viscosity fields, respectively, $\mathbf{g} = -g\mathbf{j}$ is the gravity acceleration, and $\boldsymbol{\sigma}$ is the stress tensor. The surface tension $\sigma$ is taken to be constant and $\delta$ is a three-dimensional delta function constructed by repeated multiplication of one-dimensional delta functions, identifying the interface location. $\kappa_f$ is twice the mean curvature, $\mathbf{n}_f$ is a unit vector normal to the front, $\mathbf{x}$ is the point at which the equation is evaluated, and $\mathbf{x}_f$ is the position of the interface. The subscript $F$ on the integral sign denotes integration over the front. The system is treated as two-phase immiscible fluid flow and the model is based on the one-fluid formulation, where the governing equations for the polymer and the air are solved together for the whole domain. In order to distinguish the different phases, a marker function $I(\mathbf{x})$, that is 0 inside the polymer and 1 in the air, is constructed from the position of the interface. Hence, in the whole domain a generalized material property $\phi$ is given as:

$$\phi(\mathbf{x}) = \phi_p(\mathbf{x}) + [\phi_a(\mathbf{x}) - \phi_p(\mathbf{x})]I(\mathbf{x}). \quad (2)$$

In order to consider the effect of buoyancy due to differences in temperature, the Boussinesq approximation is introduced into the gravity term in the momentum equations and density of the air is $\rho_a(\mathbf{x}) = \rho_a(1 - \alpha \Delta T)$. Specifically, from the ideal gas state equation, $pv = nRT$, as well as a constant-pressure assumption, the thermal expansion coefficient can be easily derived as $\alpha = 1/T_{ref}$.

The flow is assumed to be incompressible, so the conservation of mass equation is

$$\nabla \cdot u = \dot{Q}\delta(\mathbf{x} - \mathbf{x}_S) \qquad (3)$$

where $\dot{Q}$ is the volume source modeling the nozzle, located at $x_S$, in a small constant volume, a sphere with a radius $R < 0.25D$, and tracking a prescribed trajectory. For the low Reynolds numbers used here an approximation using a moving source to model the injection is likely to be reasonable. However, the nozzle model could be imporved by adding a dipole source and higher order terms for both the mass and the momentum equations. The temperature of both the polymer and the air is found by solving the energy equation

$$\frac{\partial \rho c_p T}{\partial t} + \nabla \cdot \rho c_p \mathbf{u} T = \nabla \cdot k \nabla T + \rho c_p T_{inj} \dot{Q}\delta(\mathbf{x} - \mathbf{x}_S). \qquad (4)$$

where the last term on the right side is the heat source due to the addition of the hot molten polymer. For the simulations presented here we assume that the polymer melt can be approximated as a Newtonian fluid, so the stress tensor is taken to be $\boldsymbol{\sigma} = \mu(\nabla \mathbf{u} + \nabla \mathbf{u}^T) = 2\mu\mathbf{D}$, where $\mathbf{D}$ is the rate of deformation tensor. The viscosity of the air is taken to be constant and the viscosity of the polymer to depend on the temperature and shear rate. In principle the viscosity of the polymer melt also depends on pressure and the flow may be viscoelastic. Once the hot melt exits the nozzle the pressure is close to the ambient pressure and does not change much. Thus pressure effects are ignored here. Volume changes as the material cools down and rigidifies are also ignored. Given a specific volume as a function of temperature (and pressure, although we expect that the pressure dependency is not important for FDM), the volume source terms are easily added to the computational model.

The full physical behavior of a polymer used in FDM for the whole range of the operational temperature is very complicated, and phase change, chemical reactions, and crystallization can have great impact on the properties of the final product. For a typical material used in FDM, such as PLA, three different states, fully solid, high elastic and liquid, are present as the temperature increases. The polymer leaving the nozzle is in the liquid state at a temperature above the solidus point, typically about 160°C for PLA. When the temperature decreases to the solidus point the viscosity increases dramatically. As the temperature drops further down to the glass transition point, about 60°C, the viscosity becomes very large, making the polymer rigid. In this transition stage the dynamic behavior may also be influenced by viscoelastic effects. Additionally, the enthalpy and volume of the injected material will change with its microscopic organization when the temperature crosses a phase-transition point.

Here viscoelastic effects are neglected and the Cross-WLF model (Bilovol, 2003) is applied for the viscosity:

$$\mu = \frac{\mu_0}{1+(\frac{\mu_0 \dot{\gamma}}{\tau^*})^{(1-n)}}. \qquad (5)$$

In Equation (5), $\mu_0$ is the zero-shear rate viscosity, $\dot{\gamma}$ is the shear rate, calculated by $\dot{\gamma} = \sqrt{2\mathbf{D}:\mathbf{D}}$, $\tau^*$ is the critical stress at the transition to shear thinning and $n$ is a data-fitted power law index. We use $\tau^* = 1.00861 \times 10^{+5}$ and $n = 0.2500$. The zero-shear rate viscosity is given by:

$$\mu_0 = D_1 exp\left[-\frac{A_1(T-T^*)}{A_2+D_3 p+(T-T^*)}\right], \qquad (6)$$

where the reference temperature $T^* = D_2 + D_3 p$ is pressure depended. Here, $D_1 = 3.31719 \times$

$10^{+9}$, $D_2 = 100$, $D_3 = 0$, $A_1 = 20.194$ and $A_2 = 51.6$ (Moldow Plastic Labs, 2007) are data-fitted constants. Since pressure effect is ignored, the final viscosity of the melt depends only on the temperature and the shear rate, as shown in Figure 2. Since the value of the viscosity at low temperature and shear rate is very high, a value of $5000 Pas$ is given as the maximum viscosity, which is large enough to immobilize the injected material.

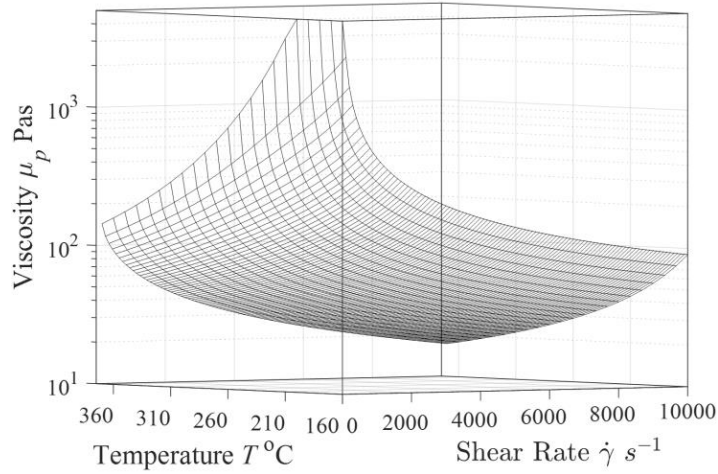

Figure 2: Temperature and shear rate dependent viscosity for the melt.

The computations are done using a numerical method introduced in Unverdi and Tryggvason, 1992, to simulate the evolution of flows of two or more immiscible fluids. The Navier-Stokes equations are solved for the fluid velocity using a projection method and finite volume numerical approximation on a regular structured and staggered grid and the interface between the different fluids is tracked using triangular grid constructed from marker points. The velocity of the marker points is interpolated from the fixed grid and the points are moved with the fluid. The new interface location is used to generate a marker function on the fluid grid, and then the material parameters, such as the density, viscosity, and thermal conductivity of the fluid are set using Equation (2). The marker points are also used to find the surface tension, which is transferred to the fluid grid as a volume force. As the molten polymer is injected, the interface area increases and new points must be inserted as needed. For a detailed description of the original method, and various validation and accuracy tests, see Tryggvason et al., 2001, 2011. For the present problem, the large ratio of material parameters between the air and the melt, especially for the viscosity and density, can pose challenges for the original numerical method which uses explicit time integration. The ratio of viscosities is in the range $\mu_p/\mu_a = 8.70 \times 10^5 - 2.18 \times 10^8$, for the density we have density ratio of $\rho_p/\rho_a = 1378$, and for the thermal diffusivity the ratio is $\alpha_a/\alpha_p = 480$. Thus, the computations of the viscous terms in the momentum equation and the thermal diffusive term in the energy equation have been modified and solved implicitly.

The implementation of the whole algorithm is summarized below:
1. Compute the surface tension from the current interface location and transfer it to the fluid grid as a volume force, denoted by $F^n$;
2. Advect the marker points explicitly: $\mathbf{x}_f^{n+1} = \mathbf{x}_f^n + \mathbf{u}(\mathbf{x}_f)^n \Delta t$;
3. Check the interface grid and add/delete points if its quality is poor;
4. Find the marker function by integrating the gradient of the marker function (see Tryggvason et al., 2011);
5. Set the material properties $\phi^{n+1}$;

6. The momentum equations are integrated in time using a standard projection scheme, where a projected velocity ($\mathbf{u}^*$) is first found, a pressure equation is solved, and the velocity then corrected by adding the pressure gradient. To deal with the high ratio of the viscosity between the polymer and air, the viscous terms are split into two parts $\mathbf{T} = \frac{1}{2}(\mathbf{T}^n + \mathbf{T}^*)$ (Kim and Moin, 1985), where $\mathbf{T}^n$ is evaluated using the old velocity $\mathbf{u}^n$ but $\mathbf{T}^*$ using the predicted velocity $\mathbf{u}^*$. The equation for $\mathbf{u}^*$ is:

$$\frac{\rho^{n+1}\mathbf{u}^* - \rho^n \mathbf{u}^n}{\Delta t} = \frac{1}{2}\mathbf{T}^* + \frac{1}{2}\mathbf{T}^n - \mathbf{A}^n + \mathbf{S}^n, \tag{7}$$

where $\mathbf{S}^n$ and $\mathbf{A}^n$ denote the source term and the advection term, computed at the old time step. The predicted velocity is then corrected, using

$$\frac{\rho^{n+1}\mathbf{u}^{n+1} - \rho^{n+1}\mathbf{u}^*}{\Delta t} = -\nabla p'. \tag{8}$$

Since there is a difference between $\mathbf{T}^*$ and $\mathbf{T}^{n+1}$, a pressure-like term $p'$ replaces the original pressure (Kim and Moin, 1985). The steps are:

6.1. Calculate the source term in the momentum equation, $\mathbf{S}^n = \rho^n \mathbf{g} + \mathbf{F}^n$, and the advection term $\mathbf{A}^n$;
6.2. Solve Equation (7) to find the projected velocity $\mathbf{u}^*$;
6.3. Apply the mass conservation equation to derive a pressure-like from Equation (8), giving

$$\frac{\dot{Q}\delta(\mathbf{x} - \mathbf{x}_S^{n+1}) - \nabla \cdot \mathbf{u}^*}{\Delta t} = -\nabla \cdot \left(\frac{\nabla p'}{\rho^{n+1}}\right). \tag{9}$$

Here, the material is continually added into the computational domain by the specific volume rate $\dot{Q}$, and $\mathbf{x}_S$ gives the nozzle position that is moving along a prescribed trajectory. Solve Equation (9) to find $p'$;
6.4. Calculate the new velocity $\mathbf{u}^{n+1}$ using Equation (8).

7. The thermal diffusive term is approximated in the same way as the viscous term, and the discrete energy equation is:

$$\frac{\rho^{n+1} c_p^{n+1} T^{n+1} - \rho^n c_p^n T^n}{\Delta t} = \frac{1}{2} R^{n+1} + \frac{1}{2} R^n - A^n + S^n. \tag{10}$$

Here $R$, $A$ and $S$ denote the diffusive, advective and source term, respectively. Solve Equation (10) for $T^{n+1}$.

The spatial derivative is computed by centered differences and the time integration is done by a second-order predictor-corrector scheme.

The above algorithm is implemented using specific boundary conditions, which for the FDM process shown in Figure 1, are:

1. For the bottom wall: a stationary wall boundary condition is used for the velocity, $\mathbf{u}_{wall} = \mathbf{0}$; a fixed temperature $T_p = 40°C$ is set at the bottom of the plate, whose thickness is $5mm$;

2. For the lateral sides: an outflow boundary condition is applied for the velocity: $\frac{\partial \mathbf{u}}{\partial \mathbf{n}} = \mathbf{0}$, where $\mathbf{n}$ is the normal direction to the side wall. The heat exchange with the outside of the domain is taken to be thermal convection: $k\frac{\partial T}{\partial \mathbf{n}} = h(T - T_\infty)$, where $h$ is the thermal convection coefficient, $h = 20 W/m^2 K$ (Bellini, 2002);

3. For the top surface a constant pressure and the same boundary conditions for temperature as on the side walls are used.

# 3 Results

## 3.1 Application of the method

Before examining the accuracy of our method and the dependencies of the FDM process on the control parameters, we simulate a representative object to demonstrate the capabilities of the method. Figure 3 shows several frames from a simulation of the fabrication of a simple rectangular object, composed of an outer shell, one filament thick, and an infill consisting of a filament with three U-bends spaced about quarter filament diameter apart and laid down perpendicular to each other. The material parameters used here are given in Table 1. The computational domain is $6.065 \times 1.957 \times 6.065$ millimeters and resolved by a $186 \times 60 \times 186$ grid. The nozzle velocity is $U_s = 0.10 m/s$ with injection rate $\dot{Q} = 4.81 \times 10^{-8} m^3/s$, giving a nominal diameter of $D = 0.783 mm$. The temperature of the melt when it leaves the nozzle is 215 °C and the ambient air is initially at 20 °C. The prescribed trajectories for the nozzle are $0.40D$ above the bottom for the bottom filament, and $1.30D$ for the top filament.

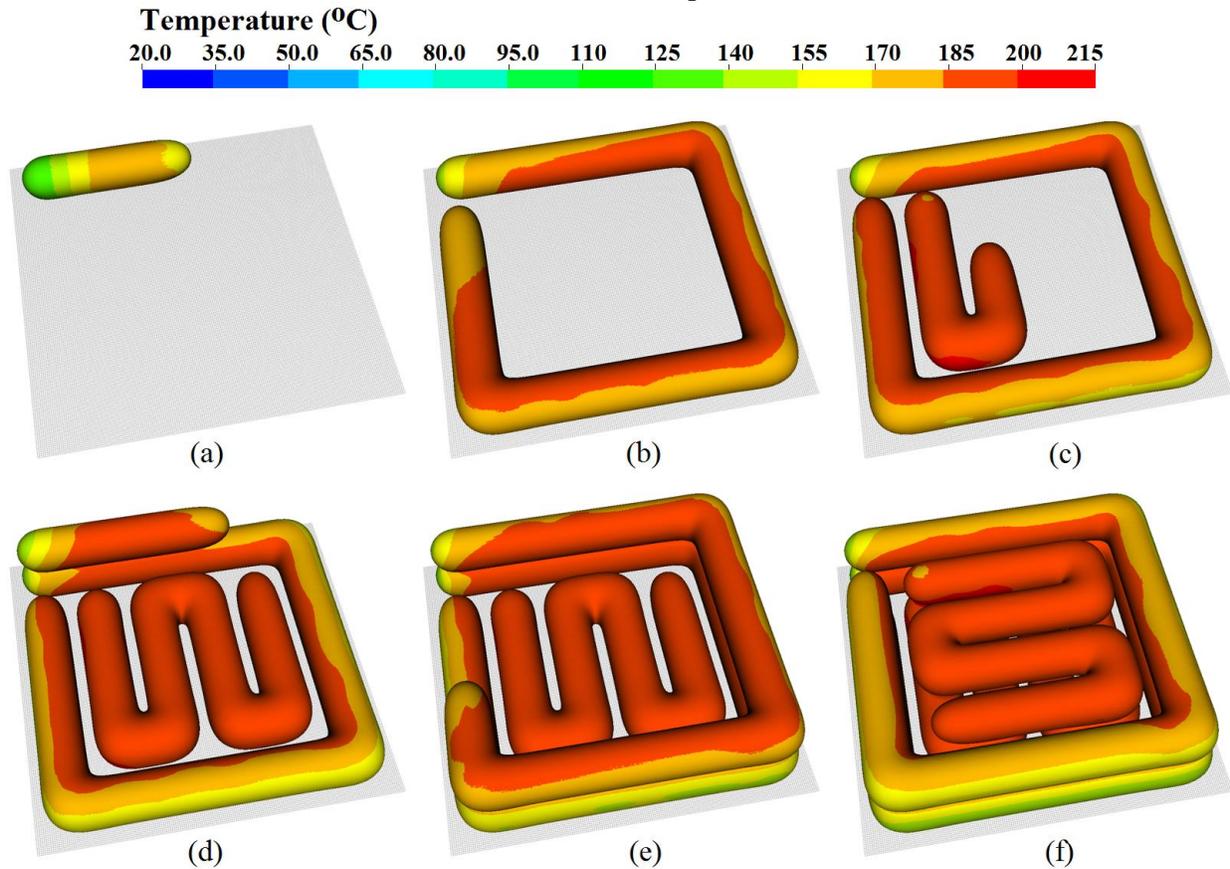

Figure 3: A few frames, (a)-(f) in time order, from a simulation of the making of a two-layer object where the filaments are laid down in parallel and perpendicular direction.

This simple rectangular object is a typical micro structure of a FDM constructed object, where filaments in adjacent layers are laid down on top of each other, either parallel as in the outer edge or in the perpendicular direction as in the interior. In the first frame the injection has just started and most of the filament is relatively hot. After the melt exits the nozzle, the decreasing temperature and the very low shear rate increase the viscosity very rapidly and thus the filament retains its shape, even thought the surface tension is fairly low. In the second frame the filament has been laid down to form an almost complete square and outside of the object continues to cool down. In the third frame, the outer square and one side of the first U-bend of the infill filament has been completed and the second side is being laid down. It can be seen that the temperature of the infill filament is little higher than the outer ones. In the fourth frame the outer filament in the second layer has been started. This process continues in the fifth frame. The injection is completed in frame six, where it clearly shows that the cooling goes gradually from outward to inward of the object, and the previously laid down filament has cooled more.

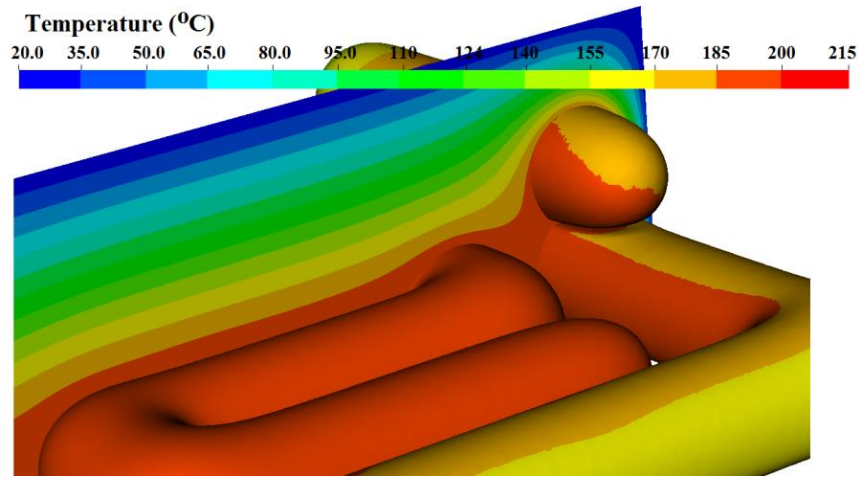

(a)

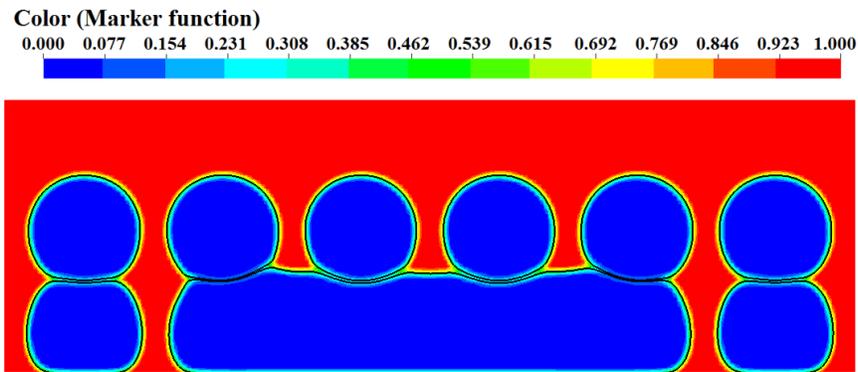

(b)

Figure 4: Detailed view. (a) The top frame shows the temperature at the surface of the filament and in one cross section. (b) The bottom frame shows the marker function, in a cross section through the middle of the object.

Figure 4, where a close-up and a cross section of the object are shown, gives a more detailed view. The top frame shows the temperature at the surface of the filament and in one cross section cutting through the newly laid down filament and the air. The bottom frame shows the marker function, identifying the filament, in one plane cutting through the final object. The temperature gradients in the top frame indicate that the hot filament has cooled down from the top and in the lateral directions by the ambient air and the top of the bottom filament has been reheated a little by the just injected hot top filament. The reheating is important in the FDM process since the reduction in viscosity allows the already laid down filament to remelt and bond with the new

filament. This increases the sticking strength between the two filaments. The cross section in the bottom frame shows that in the interior of the object the top filaments, that are laid down perpendicular to the bottom filaments, squeezes the bottom filaments, and that the outer filaments, laid down parallel on the top of each other, have a slightly flattened cross sectional area.

## 3.2 Accuracy test

The object in Figure 3 and 4 was built for demonstration purpose, using a resolution that is sufficiently fine so that the solution is likely to be fully converged, yet coarse enough to result in reasonable computer times. Below the process is examined more quantitatively, but only for a simple situation where two short filaments are laid down, one on top of each other. This simple case allows us to conduct several simulations relatively quickly and to examine the effect of the grid resolution, as well as some of the governing parameters. The computational domain is $3.913 \times 1.957 \times 1.174$ millimeters, the injection rate is $\dot{Q} = 4.81 \times 10^{-8} m^3/s$ and the speed of the nozzle is $U_s = 0.10 m/s$, giving a nominal diameter of $D = 0.783 mm$. The injection point (the source) moves along a trajectory $0.40D$ above the bottom wall for the lower filament and $1.30D$ for the top one. Other parameters are the same as listed in Table 1. In order to show the accuracy of the numerical computation, we carry out a convergence study with a benchmark grid resolution of $120 \times 60 \times 36$ grid points and three other resolutions of $60 \times 30 \times 18$, $180 \times 90 \times 54$ and $240 \times 120 \times 72$ grid points. Since the thermal diffusion rate is relatively low comparing with the nozzle velocity, the heat does not have enough time to fully diffuse to the surrounding while the nozzle is moving. Thus, to show the effect of cooling, we give the filament three seconds to cool down after each filament is finished. The filaments are shown at the end of the last cooling periode, at which time the filament ends are colder than the middle. The numerical code is fully parallel and has been run on several processors. The benchmark case takes about 30 hours on a few year old Intel processor using 36 cores and the most finely resolved case takes about three days using 84 cores. Figure 5 shows the two filaments for four levels of grid refinements. On the left the filaments and their surface temperatures are shown and the corresponding marker function in a cross section through both filaments is present on the right, at the place marked by the thin black line. The shape of the filaments is essentially the same for all the grids, but the contact area between the filaments and the temperature on the surfaces are slightly different.

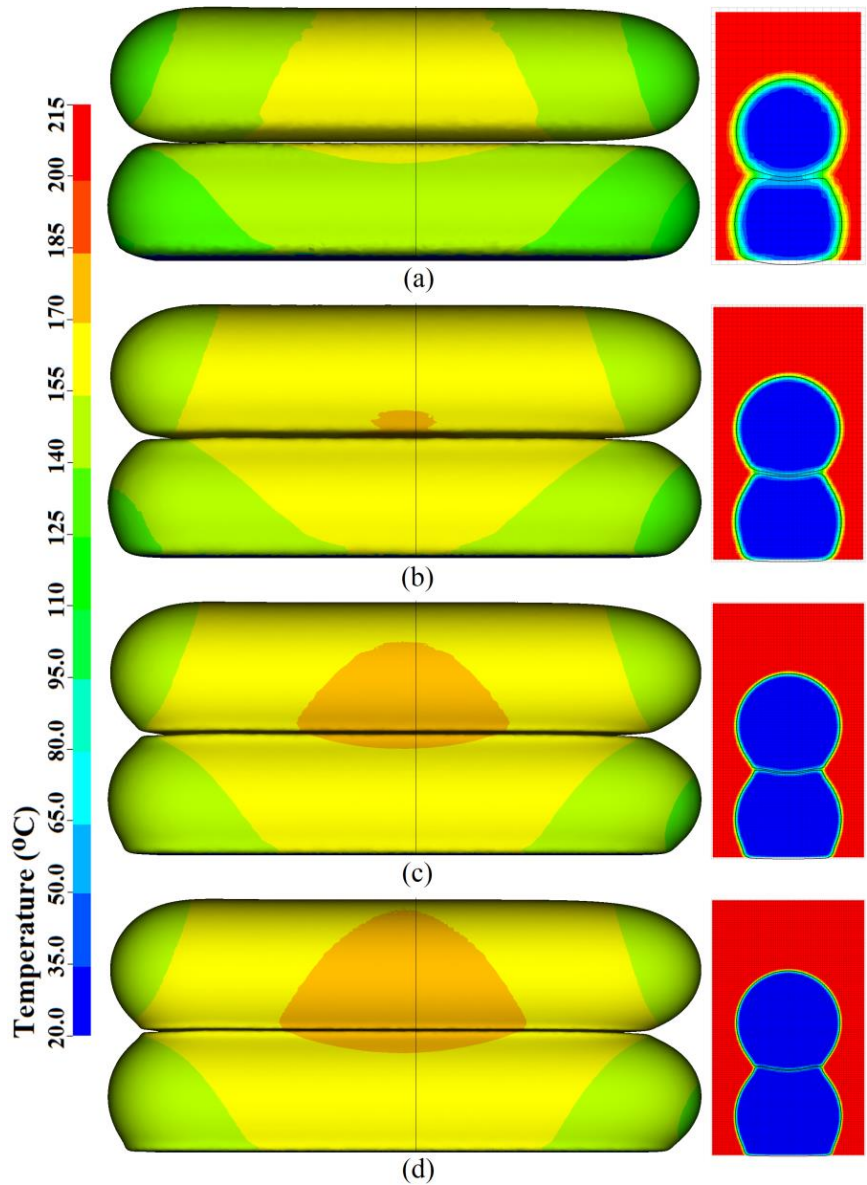

Figure 5: Two filaments deposited on top of each other, computed using four levels of grid refinement: (a) $60 \times 30 \times 18$, (b) $120 \times 60 \times 36$, (c) $180 \times 90 \times 54$ and (d) $240 \times 120 \times 72$. In the left column the filaments and the surface temperature, 3 seconds after the deposition is completed. The right column shows the marker function in the cross section indicated by the vertical black line on the left.

The contact area between the two filaments and the plate is one of the main quantities of interest (QOI), since it plays an important role in determining the strength of the resulting object. Thus, we have monitored it as we refine the grid. Since the grid resolution determines how close the interfaces are and although the filaments have been squeezed together so that the contact area is flat, for the coarsest grid they appear to be separated when looking from the side, parallel to the contact area. Due to the finite resolution, the filament surfaces will be close but touch only in the sense that the distance between them is smaller than, or of the order of, the resolution. Hence, the contact area is found using the local resolution. The contact area is computed by directly measuring the distance of two interface points on the different filaments or between a filament and the bottom wall and if the distance is a grid spacing or closer, then the fronts are identified as being is contact. Figure 6 shows the contact area found in this way at two times, first when the bottom filament has been laid down and then when the whole top filament has been added. The increase in contact area versus time is plotted in Figure 7 for four different levels of grid refinement. The local resolution measured contact area between the two filaments are essentially the same, except for on the coarsest grid, and the differences of the contact area with the bottom surface between the adjacent refined grid are smaller and smaller. The two finest grids give essentially the same results, showing that the contact area has converged. Furthermore, the contact area shows an almost linear increase that indicates that the distortion and dimensional errors of the deposited filament are very small. For a real FDM processes, the situation is more complicate because the filaments diffuse into each other, bonding together. In our current model, the diffusion and wetting phenomenon are not included. However, it should be possible to include those using the analytic sintering model proposed by Pokluda et al., 1997.

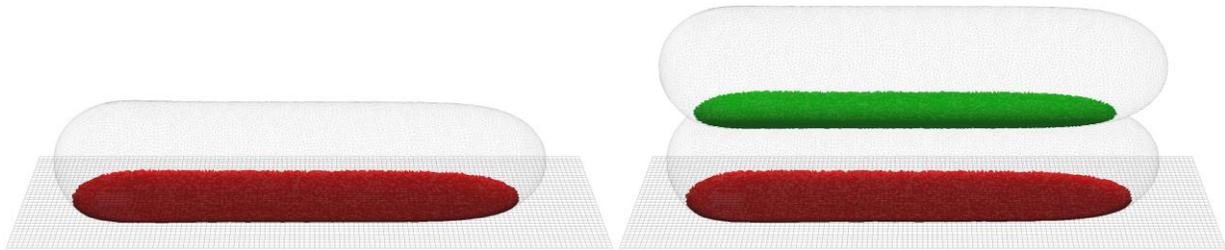

Figure 6: Contact area at two times. The red area is the contact area of the bottom filament with the wall, and the green is the contact area between the filaments.

The volumetric average temperature of the two filaments is another QOI that allows us to check how the solution to the energy equation converges and also gives more detail information about how the molten polymer cools down. Figure 8 shows the average temperature versus time. The differences among the four levels of resolution indicate that the temperature has converged. During each injecting stage, the average temperature increases and in each cooling stage it decreases. It is easy to understand how the average temperature increases and more so during injection of the top filament. Since the injection of the melt is relatively fast, the average temperature mostly represents the proportion of the hot polymer that just left the nozzle. However, for the cooling process, we see that the rate of cooling for the second stage is lower than that of the first, since the ambient air is now hotter. In addition to checking the grid convergence, we have also compared the average temperature along the filament to the simple one-dimensional analytical model for the average temperature during the injection presented by Turner, 2014, and although the setup is not exactly the same, the results are qualitatively similar.

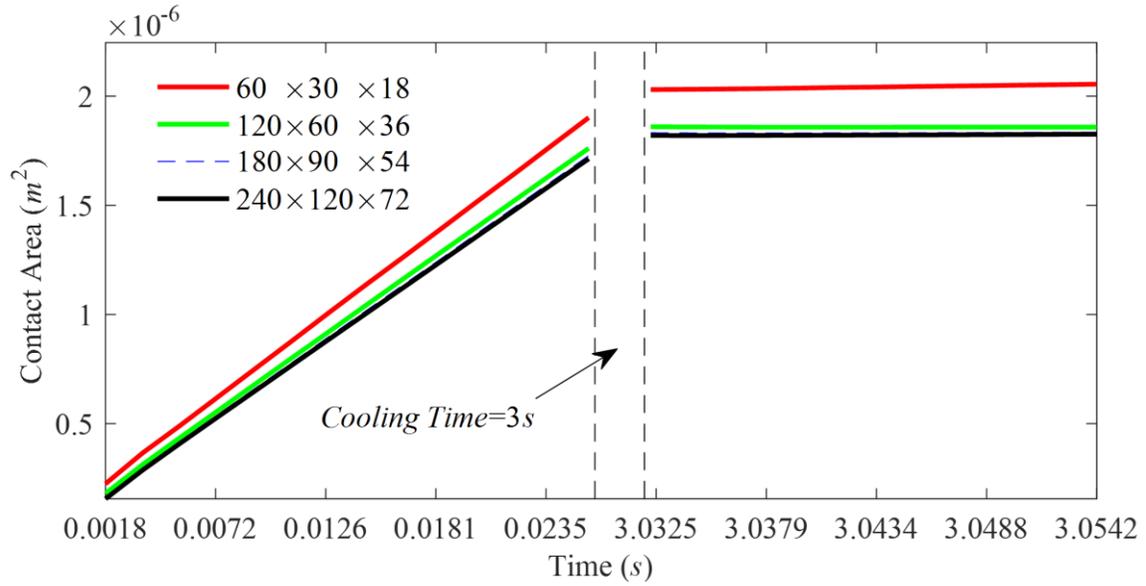

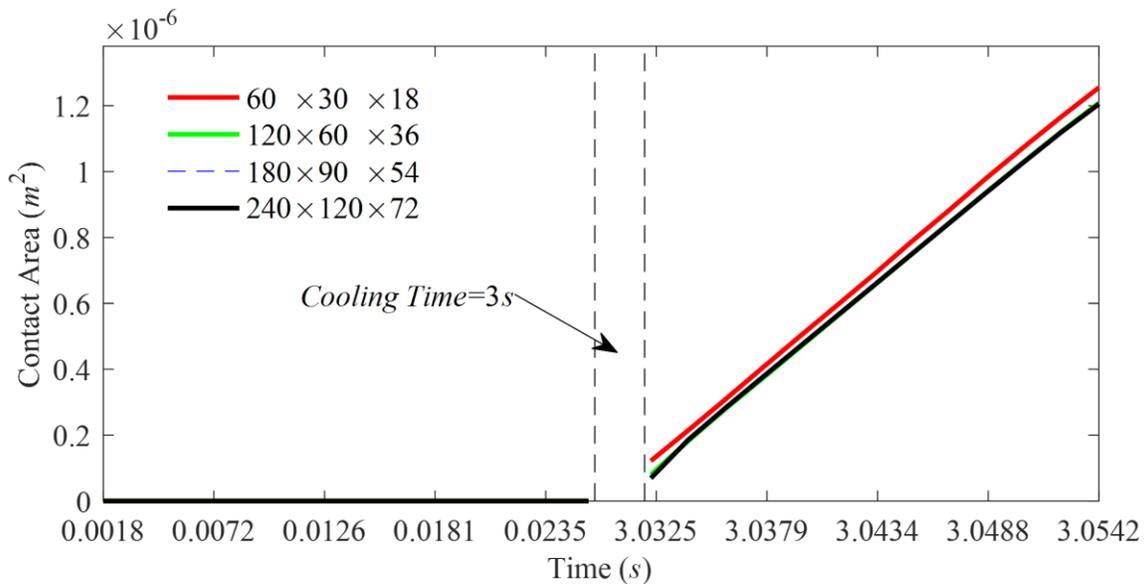

Figure 7: Contact area versus time for four different resolutions. (a) Contact area with the bottom surface. (b) Contact area between the two filaments.

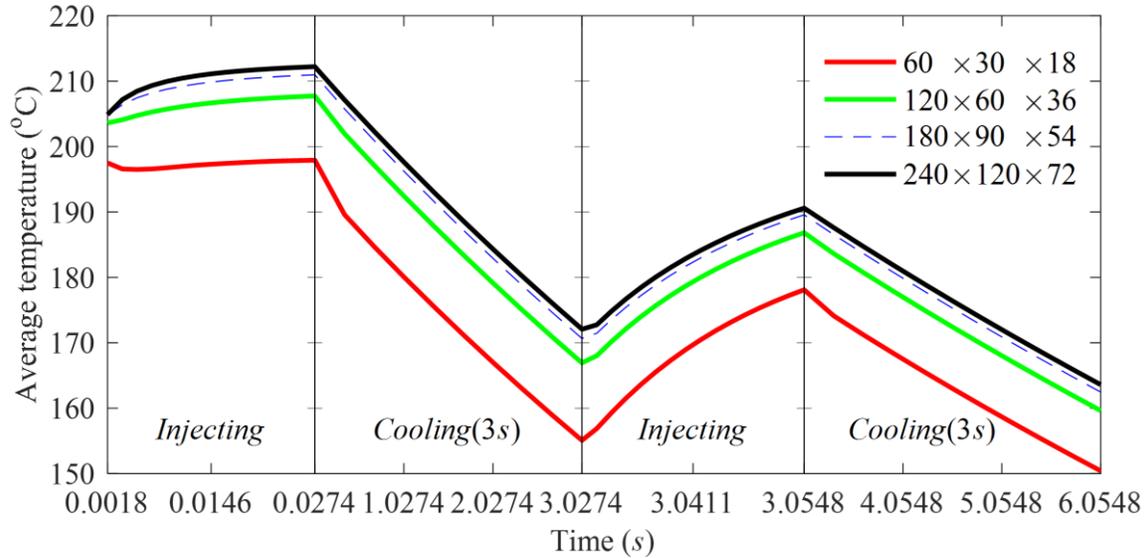

Figure 8: Average temperature of the polymer versus time. The printing sequence consists of injecting, cooling, injecting and then cooling.

### 3.3 Parametric effects

The numerical method developed here allows us to examine in detail the role of the governing parameters and here the effect of changing the nozzle velocity and the injection temperature are present. These are the two key control variables, in addition to the nozzle trajectory. In each case we do three simulations, where a single variable is varied and all the others are kept the same. For the nozzle-velocity study, the bottom filament is deposited with the same velocity, $0.10 m/s$, but the deposition velocity of the top filament is changed. Figure 9 shows two straight filaments where the top filament is laid down with a nozzle speed of $0.05 m/s$, $0.10 m/s$ and $0.15 m/s$. The figure shows that higher nozzle speed results in a smaller filament diameter, as expected, showing that the thickness of each layer can be adjusted by controlling of nozzle speed. The results indicate that the dimension of a filament is sensitive to speed fluctuations. Furthermore, dimensional errors can also be caused by starting, accelerating, decelerating and stopping of the nozzle (Turner and Gold, 2015).

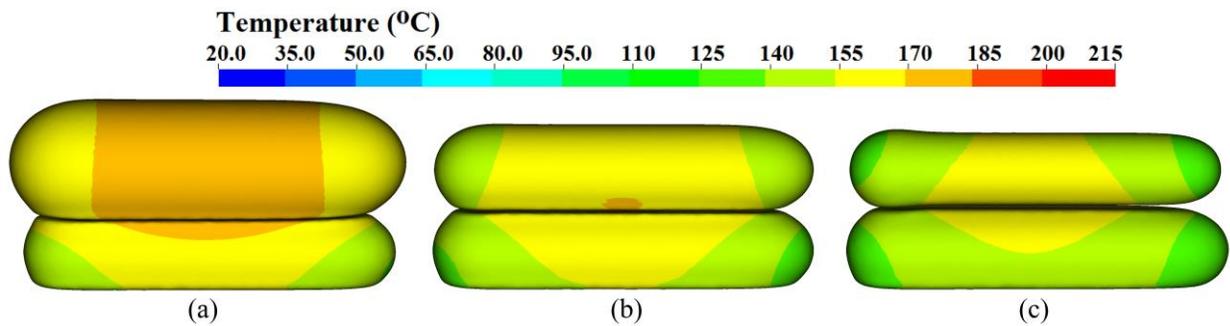

Figure 9: Two filaments deposited with different nozzle velocity, (a) $0.05 m/s$, (b) $0.10 m/s$ and (c) $0.15 m/s$.

The effect of changing the injection temperature is perhaps not as intuitive as changing the nozzle velocity, but it strongly impacts how much the bottom filament is reheated. The temperature obtained using three different injection temperatures of 215 °C, 295 °C and 375 °C are shown in the top six frames of Figure 10. While the filament shapes are essentially the same, the temperatures are different, as can be seen from the front of the filaments. The six bottom frames, at the same time as the correspongding top ones, show the temperature distribution and the filament profiles in the cross section indicated by the vertical black plane in the top frames. The temperature gradient clearly shows that the bottom filament is reheated due to heat conducted from the hot top one and the reheated area is larger for higher injection temperature. Inside the domain, the heat is added by the injected molten polymer and transferred away by conducting and convecting. For the thermal convection in the air, the effect of the buoyancy due to temperature gradient is included. Figure 11 shows the buoyancy-induced velocity field, where the hot air near the filament is rising but the cold air is flowing down from the top, and a vortex is formed by the path of the hot and cold air on the side of the hot filament.

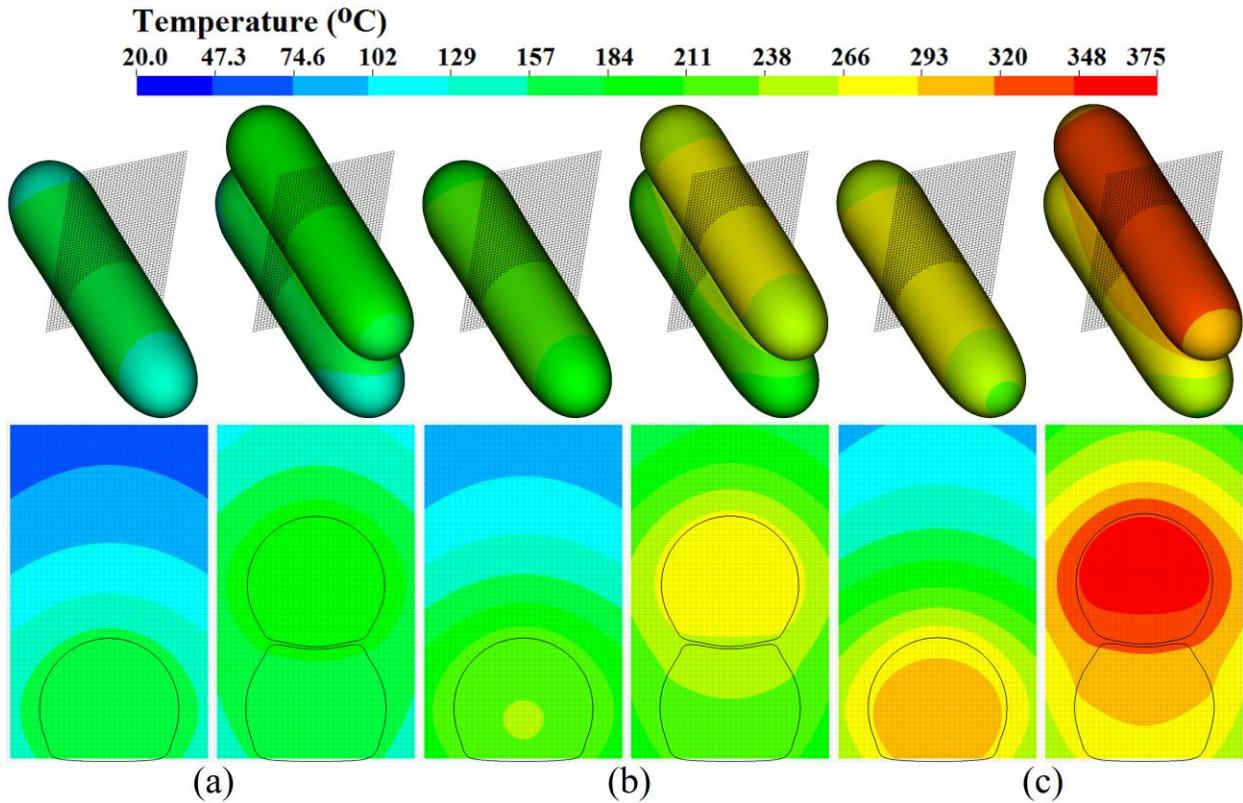

Figure 10: Two filaments deposited with different injection temperatures, (a) 215 °C, (b) 295 °C and (c) 375 °C. For each injection temperature, four frames at two times are shown. The left one in the top row is the filament and the surface temperature when the bottom filament was laid down and naturally cooled for three seconds, the right one is at the time when the deposition is completed, and the two in the bottom row show the spatial temperature and filament profile, in the cross section indicated by the vertical black surface in the top frames, at the corresponding times.

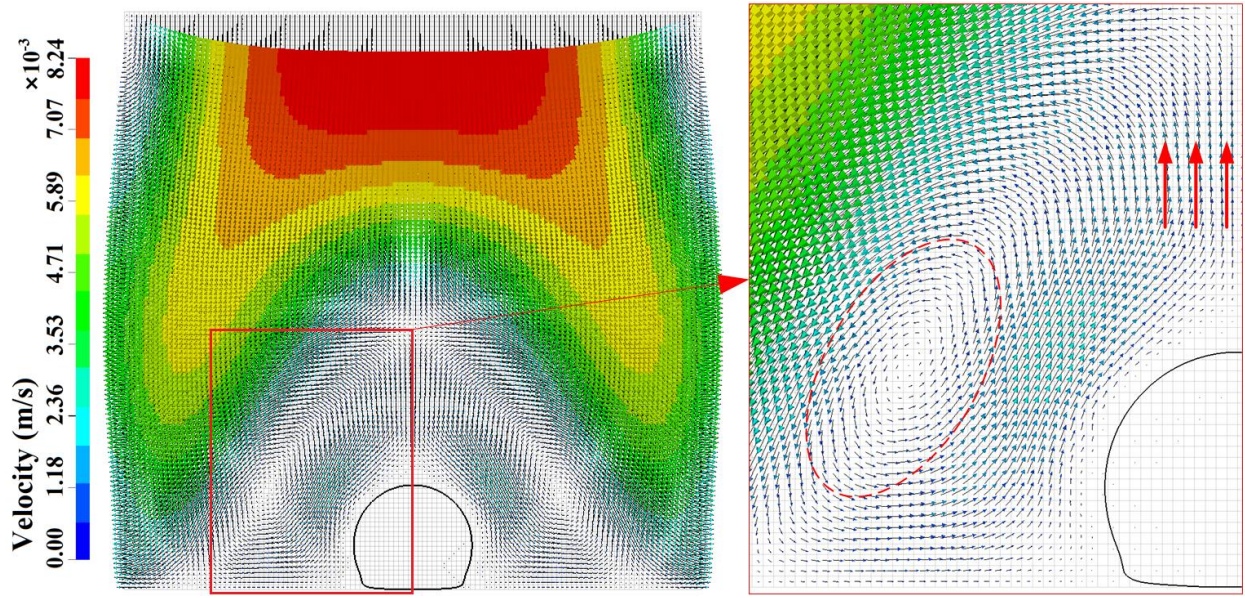

Figure 11: The velocity field induced by buoyancy. The red circle in the frame on the right marks a buoyancy created vortex.

To examine the reheat we show, in Figure 12, the temperature along a perpendicular line of variable length through the center of the bottom filament (the black plane in the top frames in Figure 10), versus time, for the three different injection-temperatures. The temperature record starts at a time when the top filament has just being deposited. As the top filament is approaching, the top of the bottom filament is pushed down, as the sudden drop in the top profile shows. The temperature and its diffusive depth at the top increases. As the nozzle moves past the line that we are monitoring the temperature gradually decreases but the deformation remains. The red contour line shows where $\Delta T = 10°C$. It is clear that the temperature change is larger and more persistent for higher injection temperature. However, the situation is different at the bottom, where the temperature drops faster for higher injection temperature due to the higher cooling rate. The frames on the right hand side are zoom-ins of the region marked by the red rectangles in the frames on the left, showing in more detail what happens as the top filament arrives. A more quantitative picture of the reheat process is shown in Figure 13a, where the location of the $\Delta T = 10°C$ contour line is plotted versus time. Here it is clear that the higher injection temperature leads to deeper and longer lasting reheat zone. The deformation of the bottom filament, plotted versus time in Figure 13b, shows a slightly smaller deformation for the higher injection temperature, at least for the parameters examined here. The contact area between the filaments (not shown) is similarly found to be slightly smaller for the higher injection temperature.

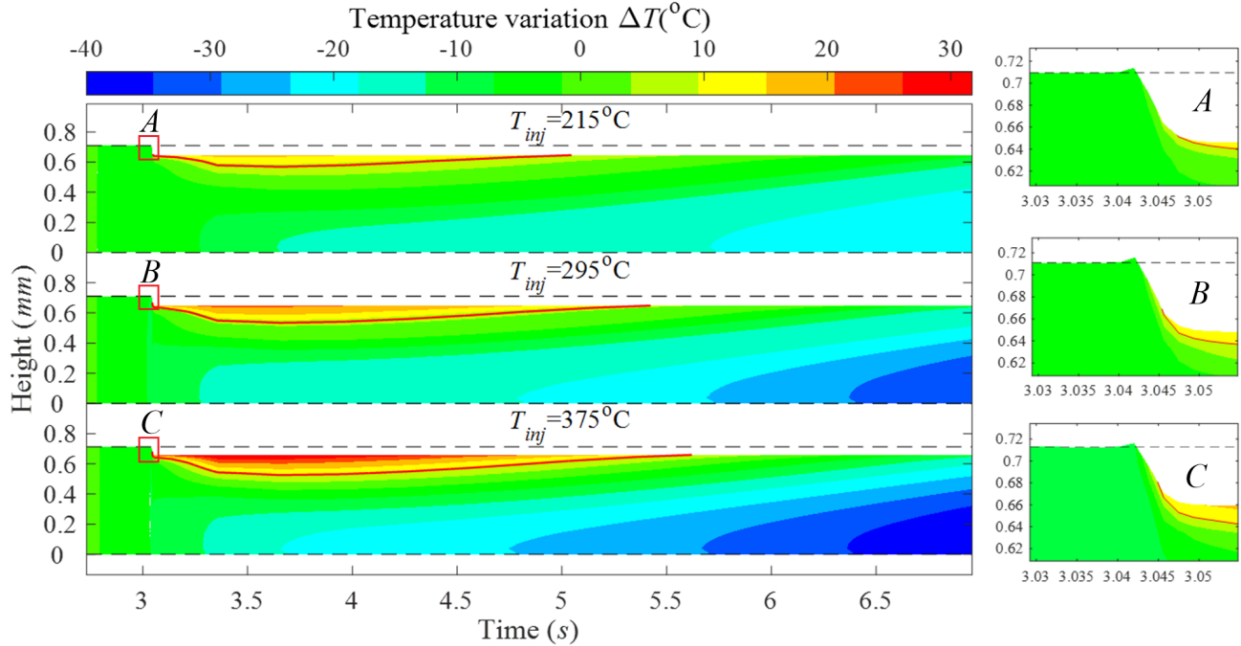

Figure 12: The evolution of the temperature variation and deformation along a perpendicular line inside the bottom filament for three different injection temperatures. The left frames are for 215 °C, 295 °C and 375 °C, respectively. The right frames are a close-up of the left frames indicated by a red rectangle and marked with the same letters, A, B and C. The red line in each frame identifies a contour line that divides the reheated zone from the part which continues to cool down.

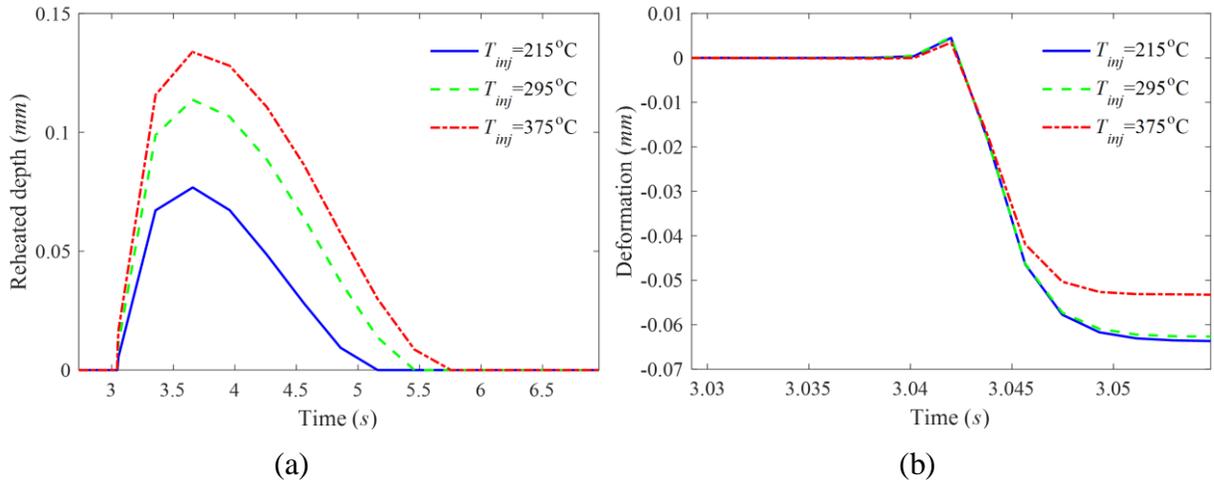

(a)                            (b)

Figure 13: (a) Reheated depth, and (b) contact area, versus time, for three different injection temperatures.

During the initial injection we sometimes saw a very high viscosity "skin" on the filament as the outer edge of the filament cooled down. However, a detailed study of this phenomenon using a one-dimensional model suggests that for the parameters used here the air heats up very quickly and the film disappears. Thus, we have not attempted to resolve it. For situations where is must be included, it is likely that a multiscale approach along the lines used by Aboulhasanzadeh et al. 2012 could be used.

# 4 Conclusion

The controlled deposition of material in a prescribed and precise manner is opening up fundamentally new opportunities in manufacturing, ranging from the inexpensive creation of personalized objects at home to very sophisticated parts of jet engines operating under extreme conditions. Several such additive manufacturing processes now exist, for a wide range of materials. Of those, fusion deposition modeling (FDM), where a stream of molten thermoplastics is ejected from a hot nozzle to build an artifact layer by layer, is perhaps the most popular one, particularly when low-cost is important. Indeed, most of the cheap personal 3D printers that have recently become available use FDM. As anyone who has used a 3D printer can attest to, the quality of the product depends on the operating parameters and these generally need to be determined by trial and error. Predicting how the final product depends on the operation parameters is a complex problem that involves unsteady fluid flow, a free boundary, heat transfer and solidification.

Over the last decade and a half significant progress has been made in the development of advanced computational tools for simulations of similar processes, but until now those have not been adopted for modeling of FDM. Here we have presented a mathematical model and a numerical method to simulate FDM, including the effect of the various operational variables. The approach is based on a method that has already been used successfully for simulations of a wide range of multiphase flows problems, including solidification. Here it has been extended to follow the injection and cooling down of a polymer. The major changes are the addition of a source to model the nozzle, the implicit treatment of the diffusion terms in the momentum and the energy equation, and the addition of a temperature and shear dependent viscosity. In the present paper we have focused on the fluid flow only, but the basic approach can also be used for solidification and the emergence of residual stresses. We use a simple volume source to represent the nozzle, but more complex nozzles can be modeled using higher order terms such as dipoles and quadrapoles. A momentum source can also be added, but that is unlikely to be needed given the low Reynolds number in the melt.

The availability of sophisticated and accurate numerical tools is likely to not only transform how the operating parameters for 3D printers are selected but, more importantly, greatly aid in the development of more sophisticated processes such as those involving multiple nozzles, co-injection of different materials, and the use of new materials. We conclude by making the rather obvious point that we do not envision that a computationally demanding methodology as the one presented here will be used for routine planning of FDM construction of objects. Rather, we envision that full simulations will be used to examine new and complex processes such as those involving multiple nozzles, co-injection of different materials, and the use of new materials. Such simulations should also provide the foundations for the construction of high fidelity reduced order models, that can be used on a more routine basis to plan how to build objects using FDM.